\documentclass[twocolumn,showpacs,preprintnumbers,amsmath,amssymb]{revtex4}%
\usepackage{graphicx}
\usepackage{dcolumn}
\usepackage{bm}
\usepackage{amsmath}
\usepackage{amsfonts}
\usepackage{amssymb}

\begin{document}

\title{Cumulative identical spin rotation effects in collisionless trapped atomic gases}
\author{F. Pi\'echon,$^1$ J.N.\ Fuchs,$^1$ and F.\ Lalo\"{e}$^2$}
\affiliation{$^1$Laboratoire de Physique des Solides, CNRS UMR 8502, Univ. Paris-Sud, F-91405 Orsay, France}
\affiliation{$^2$Laboratoire Kastler Brossel, ENS, UPMC, CNRS; 24 rue Lhomond, F-75005 Paris, France}
\date{\today}

\begin{abstract}
We discuss the strong spin segregation in a dilute trapped Fermi gas
recently observed by Du \emph{et al.} with ``anomalous'' large time
scale and amplitude. In a collisionless regime, the atoms oscillate
rapidly in the trap and average the inhomogeneous external field in
an energy dependent way, which controls their transverse spin
precession frequency. During interactions between atoms with
different spin directions, the identical spin rotation effect (ISRE)
transfers atoms to the up or down spin state, depending on their
motional energy. Since low energy atoms are closer to the center of
the trap than high energy atoms, the final outcome is a strong
correlation between spins and positions.
\end{abstract}

\maketitle

Spin waves in dilute gases were predicted at the beginning of the eighties
\cite{Bashkin, Laloe} and confirmed experimentally soon after
\cite{H,He,Gully}. They can be understood in two equivalent ways,
either as a consequence of spin mean field \cite{Bashkin}, or in
more microscopic terms as the cumulative result of the identical
spin rotation effect (ISRE) - an effect taking place during binary
collisions between identical atoms \cite{Laloe}. Similarly, the
Faraday effect is a rotation of the spin of photons that can be
seen, either as a consequence of a macroscopic index of refraction,
or as resulting from the accumulation of microscopic forward scattering
events between photons and atoms.

Experiments with ultracold atomic gases have renewed the interest in
spin waves. In 2002, a group at JILA \cite{Cornell}
showed that, in an trapped ultracold atomic gas of bosons, the ISRE
can result in a spontaneous spatial \textquotedblleft
segregation\textquotedblright\ of two atomic internal states
$\left\vert 1\right\rangle $ and $\left\vert 2\right\rangle $
(equivalent to a pseudo-spin $1/2$). 
Several groups then proposed a theoretical explanation of
these observations, using either one-dimensional spin $1/2$
hydrodynamic \cite{Levitov} or kinetic \cite{Fuchs, Williams} equations.
More recently, Du et al.\ from Duke University \cite{Thomas} did an
experiment that explores the properties of spin waves in quantum
gases of fermions ($^6$Li) in the collisionless Knudsen regime,
while most previous experiments were performed in the hydrodynamic
regime (see nevertheless \cite{Bigelow}). The spin segregation they
observe is a hundred times larger and a hundred times slower than
would be predicted by hydrodynamic theory. They call this
spectacular effect \textquotedblleft anomalous spin
segregation\textquotedblright\ and suggest that its explanation may
require \textquotedblleft a modification of spin wave theory\ or
possibly a new mechanism\textquotedblright\ for fermions.

The purpose of this letter is twofold. First we argue that no
modification of spin wave theory is necessary to understand the
experiment; the physical mechanism behind the observations is the
usual ISRE. The difference between bosons and fermions is not
essential; what is important is the collisionless regime. 
Second, we discuss why this
collisionless regime, combined with the presence of a trap, gives
access to unexpected and interesting new physics. In fact, one can
observe the ISRE almost as in an ideal experiment, where a single
spin polarized atom is sent through a target of a gas polarized in
another direction, and where the spin direction of the outgoing atom
is measured. Moreover, when the trap potential sends back the atom
through the same target, the rotations of the spins are additive
(the symmetry of ISRE is the same as that of the Faraday effect). As
a result, one reaches situations where the spins of the atoms are
correlated to their motional energies in the trap, instead of their
positions; these situations are inaccessible in the hydrodynamic
regime, where the value of the spin current is determined locally by
a thermal average over many collisions.

We begin with a comparison between the JILA and Duke experiments. In
the latter, the peak density $n(0)$ and the typical scattering
length $ a_{12}$ are respectively 15 and 20 times smaller than in
the JILA experiment, making the diluteness factor $1/n(0) a_{12}^3$
of the gas $10^5$ times larger. The differences can be expressed in
terms of $5$ relevant time scales: (1) the radial trap period
$2\pi/\omega_{rad}$; (2) the axial trap period $2\pi/\omega$; (3)
the average time between (lateral) collisions $\tau \sim (4\pi
a_{12}^2 n(0)v_T)^{-1}$, where $v_T\equiv\sqrt{ k_B T/m}$ is the
thermal velocity, $T$ the temperature and $m$ the atomic mass; (4)
the typical spin precession period in the external magnetic field
$2\pi/ |\Omega_0|$; and (5) the typical precession period in the
spin mean-field $\tau_{fwd} \sim m/\hbar |a_{12}| n(0)$. These time
scales [in ascending order for the Duke exp.] are compared in the
table:
\begin{center}
\begin{tabular}{|c|c|c|c|c|c|}
\hline
Time scales [in ms] & $2\pi/\omega_{rad}$ & $2\pi/\omega$ & $2\pi/|\Omega_0|$ & $\tau_{fwd}$ & $\tau$\\
\hline Duke exp. \cite{Thomas} & 0.2 & 6.9 & 100 & 300 & 5000\\
\hline JILA exp. \cite{Cornell} & 4 & 143 & 170 & 14 & 10\\
\hline
\end{tabular}
\end{center}
where $a_{12} \sim -5a_0$ and $\Omega_0/2\pi \approx -10$ Hz are
typical values for the Duke exp. \footnote{In the Duke experiment,
the scattering length $a_{12}$ could be tuned, but the precession
frequency $\Omega_0$ was fixed; in the JILA experiment, the reverse
was true.}. The ISRE is then strong: $\tau/\tau_{fwd}\sim \hbar/(m
v_T |a_{12}|)\gg 1$. In the JILA experiment, an hydrodynamic
description was qualitatively correct because of the relatively
small value of $\omega \tau \leq 1$. In the Duke experiment $\omega
\tau \sim 4500$: the gas is so dilute that its dynamics is well
described in the collisionless limit (lateral collisions are
ignored). On average, an atom oscillates 700 times in the axial trap
between two collisions.

We base our analysis on the formalism used in \cite{Laloe} and
\cite{Fuchs}, which gives a general frame for the study of non
degenerate quantum gases. It distinguishes two effects of the
interactions: mean field effects (forward scattering in collisions)
and ``real'' collisions (lateral scattering), both with a full
treatment of the effects of spin polarization and statistics. This
leads to a kinetic equation with arbitrary position and momentum
dependence. In the hydrodynamic limit, the system remains close to
local equilibrium, momenta can be integrated out, and the spin
current is simply proportional to the local gradient of
magnetization. But we do not make such an assumption and keep a full
dependence on the variables.

Consider a cigar-shaped harmonic trap with axial ($x$ direction)
frequency $\omega/2\pi$ and radial ($y$ and $z$ directions)
frequencies $\omega_{rad}/2\pi$. As $\omega\ll \omega_{rad}$, an
effective 1D description of the dynamics is possible. The peak
density (per unit volume) $n(0)$ is related to the total number of
atoms by $N=n(0)(2\pi)^{3/2}x_T r_\perp^2$, where $x_T\equiv \sqrt{
k_B T/m\omega^2}$ (resp. $r_\perp \equiv \sqrt{ k_B
T/m\omega_{rad}^2}$) is the characteristic atomic cloud size in the
axial (resp. radial) direction. The two internal states $|1\rangle$
and $|2\rangle$ are treated as an effective spin $1/2$. The three
relevant scattering lengths are $a_{12}$, $a_{11}$ and $a_{22}$; the
corresponding coupling constants are $g_{ij}\equiv 4\pi \hbar^2
a_{ij}/m$. In spin space,  $\mathbf{e}_\parallel$ is the
longitudinal unit vector, while $\mathbf{e}_{\perp1}$,
$\mathbf{e}_{\perp2}$ correspond to the transverse directions. The
spin couples to an external effective magnetic field
$\Omega(x)\mathbf{e}_\parallel$, with an inhomogeneity along $x$
characterized by a curvature $\Omega''(0)$. Therefore, the magnetic
field is taken to be $\Omega(x)\approx \Omega_0x^2/x_T^2$, where
$\Omega_0 \equiv \Omega''(0)x_T^2/2$ is a characteristic spin
precession frequency \footnote{The relation between $\Omega_0$ and
$\delta \Omega$ defined in \cite{Fuchs} is
$\Omega''(0)=2\Omega_0/x_T^2=\delta \Omega/x_T^2$. By definition
$\hbar \Omega\equiv V_2-V_1$, see \cite{Fuchs}.}.

The kinetic equation for a spin $1/2$ non-degenerate atomic
gas (fermions or bosons) is written in terms of a $2\times 2$ operator $\hat{\rho}$
obtained by a Wigner transform of the density operator:
\begin{equation}
\hat{\rho}(x,p,t)=\frac{1}{2}\left[f(x,p,t)\hat{I}
+\mathbf{M}(x,p,t)\cdot
\hat{\mathbf{\sigma}}\right]
\end{equation}
Equivalently, we can reason in terms of the phase space density $f$ and spin
density $\mathbf{M}$. In the Duke experiment, the gas is
non-degenerate ($T \approx 27 \mu\mathrm{K}\gg T_F\approx
7\mu\textrm{K}$). As in \cite{Fuchs} we make some
approximations: both the density mean-field $|g_{12}|n(0)/4 \approx
h \times 2$~Hz and the Stern and Gerlach segregation energies $
\hbar |\Omega_0| \approx h \times 10$~Hz are negligible when compared
to the harmonic confining potential $m\omega^2x_T^2/2= k_B T/2
\approx h \times 280$~kHz. Since lateral collisions are a
small perturbation, we treat them in the simplest relaxation time
approximation. Here a single coupling constant $g_{12}\neq
0$  appears, instead of three, because $g_{11}=g_{22}=0$ results from the Pauli exclusion principle (collisions only occur between atoms
in different spin states). We then obtain the
equations \cite{Fuchs}:
\begin{eqnarray}
d_t f\equiv (\partial_{t}+p\partial_{x}-x\partial_{p})f &\simeq & -\frac{f-f^{eq}}{\tau}\label{kineticn} \\
d_{t}\mathbf{M}-[\Omega(x)\mathbf{e}_\parallel+\epsilon
g\mathbf{m}]\times\mathbf{M}&\simeq&
-\frac{\mathbf{M}-\mathbf{M}^{eq}}{\tau} \label{kineticm}
\end{eqnarray}
where $\epsilon=-1$ for fermions (and $+1$ for bosons); the superscript $^{eq}$
denotes local equilibrium phase space densities; the dimensionless coupling constant
$g$ is defined as $g\equiv g_{12}n(0)/2\hbar \omega$.
Dimensionless units have been used: lengths are measured in
units of $x_T$, momenta in units of $p_T\equiv m v_T$, angular
frequencies in units of $\omega$, times in
units of $1/\omega$, and phase space densities $f$ and $\mathbf{M}$ in units of $n(0)/p_T\sqrt{2\pi}$.
The density $n$ and the spin density $\mathbf{m}$ [both in units of $n(0)$] are defined as:
\begin{equation}
n(x,t)=\int \frac{dp}{\sqrt{2\pi}} f \textrm{ and }  \mathbf{m}(x,t)=\int \frac{dp}{\sqrt{2\pi}}
\mathbf{M}
\end{equation}
such that $1=\int dx \, n/\sqrt{2\pi}$ [and e.g. $\mathbf{e}_{\perp1}=\int dx \,
\mathbf{m}/\sqrt{2\pi}$ if the gas is fully polarized in the
$\mathbf{e}_{\perp1}$ direction].
The total density $n=n_1+n_2$ and the longitudinal spin density $m_\parallel=n_2-n_1$ are
related to the internal state populations, while the transverse spin density
$\mathbf{m}_\perp$ to their coherences.

\begin{figure}[h]
\includegraphics[height=6cm]{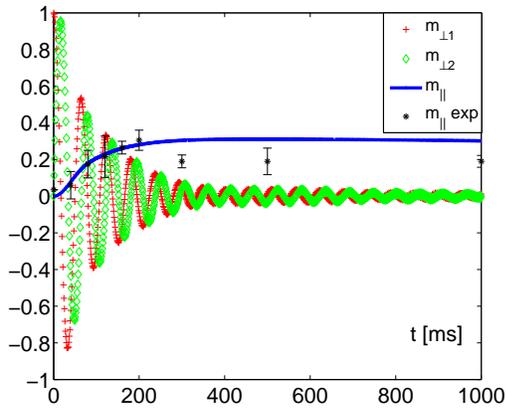} \caption{Spin density at
the center of the trap as a function of time $\mathbf{m}(x=0,t)$ [in
units of $n(0)$] for $\epsilon g =0.021$, $\Omega_0=-0.07$ and
$1/\tau=2\times 10^{-4}$. We obtain a good agreement with the
measurements of $m_\parallel=n_2-n_1$, taken from fig. 2c of
\cite{Thomas} and shown here with their error bars: both the
amplitude and the time constants of the spin segregation are well
reproduced.} \label{mzt}
\end{figure}

\begin{figure}[h]
\includegraphics[height=6cm]{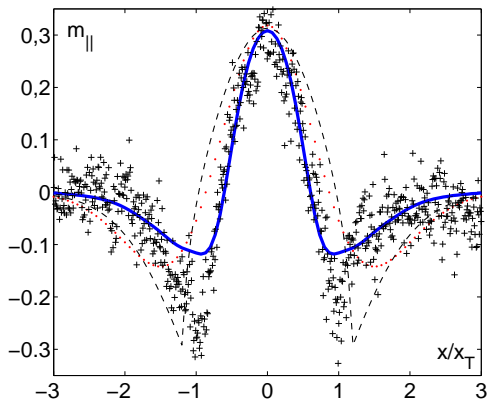}
\caption{Numerical results (full line) for the longitudinal spin
density at maximum segregation time $m_\parallel(x,t_{max})$ [in
units of $n(0)$] as a function of position; the parameters are the
same as in fig. \ref{mzt} and $x_T=210$ $\mu$m. These results are
in very good agreement with the measurements of \cite{Thomas}
(crosses) at the trap center and edges, without any adjustable
parameter. The discrepancy close to $x/x_T \approx \pm 1$ may be
attributed to a small population imbalance in the experiment ($\sim 4$\% as estimated from the area under 
the experimental profile). The
results of two approximations are also shown (dotted and dashed
lines); see text for more details.} \label{mzx}
\end{figure}

These equations are now solved numerically. Since the experiment
starts with a $\pi/2$ pulse which suddenly transfers the
spins to the transverse plane, the initial distributions are
$f_0(x,p)=\exp[-(x^2+p^2)/2]$ and
$\mathbf{M}_0(x,p)=f_0(x,p)\mathbf{e}_{\perp1}$. As $f_0$ solves the
kinetic equation, the total density does not evolve
$n(x)=\exp(-x^2/2)$ and the dynamics after the pulse can be
expressed in terms of $\mathbf{M}$ only. The 3 dimensionless
parameters appearing in the kinetic equations are taken as $\epsilon
g=0.021$ [corresponding to $a_{12}=-4.6a_0$ and $\epsilon=-1$],
$\Omega_0=-0.07$ and $1/\tau = 2\times 10^{-4}$. Fig. \ref{mzt}
compares the results with those plotted in fig. 2c of \cite{Thomas}:
both the maximum segregation time ($t_{max}\sim 300$ ms) and
amplitude ($m_\parallel(0,t_{max})/n(0) \sim 30$\%) are in good
agreement with the experimental data. Note also that the sign of the
segregation is predicted correctly. It is clear that standard theory
with a transport equation treated in the collisionless regime
explains the observations without introducing any new mechanism. The
system eventually relaxes to equilibrium on the time scale of a few
$\tau$, which can be very slow indeed because $\tau \sim 5$ s.
Calculations for other values of $a_{12}$ are in fair agreement with
the experimental data. In particular, changing the sign of $a_{12}$,
of $\Omega_0$ or of $\epsilon$ simply reverses the role of the two
internal states. When $a_{12}\approx 0$, no significant segregation
takes place. Fig. \ref{mzx} shows the longitudinal spin density
profile at saturation and a good agreement with the experiment
results \cite{Thomas}.


In order to better understand the numerical results of the previous
section, approximate analytical solutions of the linearized
[$|g|<<1$] collisionless kinetic equations are now provided. In the
absence of lateral collisions, the (axial) motional energy of an atom $E=
(x^2+p^2)/2$ [in units of $k_B T$] is a constant of the motion and
${\bf M}(x,p,t)$ conserves its norm. We therefore write
$M_{\parallel}=f_{0}\sin{\theta}$,
$M_{\perp}=M_{\perp1}+iM_{\perp2}=f_{0}\cos{\theta}e^{i\phi}$, which
define the transverse $\phi(x,p,t)$ and longitudinal $\theta(x,p,t)$
precession angles in phase space. Accordingly the initial conditions
are $\theta(x,p,0)=0$ and $\phi(x,p,0)=0$. Eq. (\ref{kineticm})
translates into coupled equations of motion for $\theta$ and $\phi$:
\begin{eqnarray}
&&d_{t}\phi=\Omega+\epsilon g(\langle\sin{\theta}\rangle-\frac{\sin{\theta}}{\cos{\theta}}\textrm{Re}\left\{ e^{i\phi}\langle\cos{\theta}e^{-i\phi}\rangle\right\} ) \label{phikinetic}\\
&&d_{t}\theta=\epsilon g \textrm{Im}\left\{ e^{i\phi}\langle\cos{\theta}e^{-i\phi}\rangle\right\} \label{thetakinetic}
\end{eqnarray}
where the momentum average is defined as $\langle A \rangle\equiv \int\frac{dp}{\sqrt{2\pi}}f_0 A$ for any function
$A(x,p,t)$. To first order in $g$, we can ignore the second term in eq. (\ref{phikinetic}) and assume $\cos{\theta}\simeq1$ in eq. (\ref{thetakinetic})
(furthermore $M_{\parallel}\simeq f_0 \theta$ and
$M_{\perp}\simeq f_0e^{i\phi}$).  These linearized equations can be solved exactly \cite{inpreparation}.
Here we just quote the results for the transverse spin precession angle $\phi(x,p,t)$:
\begin{equation}
\phi=\Omega_0 [ \frac{1}{2}\gamma_{+}(t)x^{2}+\frac{1}{2}\gamma_{-}(t)p^{2}-\gamma_{0}(t)xp ]\label{phisol}
\end{equation}
where $\gamma_{\pm}(t)\equiv t\pm\frac{\sin{2t}}{2}$, $\gamma_{0}(t) \equiv \sin^{2}t$. The
longitudinal spin density $m_{\parallel}(x,t)$ is:
\begin{equation}m_{\parallel}=\epsilon g\int_{0}^{t}dt'\frac{{\displaystyle e^{-\frac{Z}{\Lambda X}x^{2}}}}{\sqrt{X}\Lambda^{1/4}}\sin[\frac{1}{2}\arctan \frac{Y}{X}-\frac{Z}{\Lambda X}\frac{Y}{X}x^{2}]\end{equation}
where $Z(t')\equiv 1+\Omega_0^{2}(\gamma_{0}^{2}(t')+\gamma_{-}^{2}(t'))$, $Y(t,t')\equiv 2s\Omega_0(\gamma_{0}(t')c-\gamma_{-}(t')s)$,
$X(t,t')\equiv 1+s^{2}+\Omega_0^{2}(\gamma_{0}(t')s+\gamma_{-}(t')c)^{2}$,
$\Lambda(t,t')\equiv 1+(\frac{Y}{X})^{2}$,
using the shorthand notation $s\equiv \sin{(t-t')}$ and $c\equiv \cos{(t-t')}$.

Three regimes can be distinguished:
(i) for times smaller than the trap period ($t\ll1$), at a fixed position $x$,
the transverse phase difference between two atoms with momentum
$p$ and $p'$ is dominated by the $x-p$ correlation term $\gamma_0(t)$ in eq. (\ref{phisol}):
$\phi(x,p,t)-\phi(x,p',t)\simeq\Omega_0 t^{2}x(p-p')\ll1$. This is the regime leading to correlation
between velocity and transverse spin and giving:
\begin{equation}
m_{\parallel}(x,t)\approx\epsilon gn(x)^2(1-2x^{2})
2\Omega_0\frac{t^{4}}{4!}
\end{equation}
as already found in \cite{Fuchs,Williams}. For the Duke exp., this
behavior is not dominant since it occurs only at very short times
[smaller than $\sim 1$ ms]. However, for times much longer than the
trap period $t\gg1$ there is a transition to an energy dominated
regime where $\phi(x,p,t)\simeq\phi(E,t)\simeq\Omega_0 E t$, where
$\Omega_0 E$ emerges as an effective transverse precession
frequency. (ii) A second regime occurs for $1\ll t \ll1/|\Omega_0|$
where the phase difference $\phi(E,t)-\phi(E',t)\simeq\Omega_0
t(E-E')\ll1$ is small -- this is the regime of correlation between
motional energy and transverse spin. Averaging over fast axial oscillations
of small amplitude, we obtain the approximate result:
\begin{equation}
m_{\parallel}(x,t)\approx-\epsilon gn(x)^{4/3}(1-\frac{4}{3}x^{2})
\frac{\sqrt{2}}{3\sqrt{3}}\Omega_0 t^{2}
\end{equation}
This gives the dominant contribution to the spin segregation in the
Duke experiment. Note that there is a sign inversion as compared to
regime (i) and that the scaling is now $t^{2}$ instead of $t^4$
\footnote{The sign of the segregation in \cite{Thomas} (resp.
\cite{Cornell}) corresponds to that of the $t^2$ (resp. $t^4$)
law.}. The typical time for triggering the spin segregation is
$t_{trig}\sim 1/\sqrt{|g\Omega_0|}$. The dotted line of Fig. \ref{mzx} shows a plot of (10) for $t\sim
1/|\Omega_0|$. (iii) The third regime is an
ergodic regime: when $t\gg1/|\Omega_0|$, the phase difference
between high and low energy atoms becomes large
$\phi(E,t)-\phi(E',t)\simeq\Omega_0 t(E-E')\gg1$, therefore
saturation occurs. However, the saturation is not captured by the
linearized equations as it is actually related to norm conservation
and to the non-linearity of the kinetic equations
\cite{inpreparation}. By comparing the numerics to the analytical
solution of the linearized equations, we find that they agree for
the first $\sim 100$ ms.

A simple model captures the essence of the phenomenon. Assume that
we can divide the atoms into two classes, the ``hot atoms'' with
some motional energy (for instance $3 k_B T$) and the ``cold atoms''
with some lower energy (for instance $k_B T /2$). All the atoms
oscillate quickly in the trap so that the precession rate of their
spins depends on the average of the longitudinal magnetic field
along their trajectory \footnote{In a quadratic trap, the rate is
proportional to the energy since $\int dt \, \Omega[x(t)]/2\pi =
\Omega_0 E$.}. Since the hot atoms go further away from the center
of the trap, their transverse spin component rotates
at a different rate than for the cold atoms. Now, when a hot atom
crosses the cloud of cold atoms, which has a different spin
direction, its spin experiences a rotation due to the ISRE in
forward scattering. This gives to the hot atom a longitudinal spin
component that depends on the sign of the ISRE, while the cold atoms
acquire the opposite component. The effect is cumulative: each time the hot atom crosses the cloud forwards and
backwards, the rotations of its spin are additive. After some time,
a large fraction of the hot atoms is transferred towards one spin
state, a large fraction of the cold atoms to the other. A situation
where the two classes of atoms have antiparallel longitudinal spins
is stable as long as lateral collisions are ignored, since ISRE does
not affect antiparallel spin directions, and since any phase space
distribution that depends only on the energy is time invariant.
Finally, because the atoms explore a region of space that depends on
their energy, the spin state of the cold atoms is dominant at the
center of the trap, the spin state of the hot atoms at the edges.

To illustrate how this spatial separation arises from spin-energy correlations, we arbitrarily separate the atoms in two equal groups, with axial motional  energy larger or
smaller than $k_B T \ln 2$. The difference in density profile
between cold $n_<(x)$ and hot $n_>(x)$ atoms is then easily obtained
in terms of the error function:
\begin{equation}
n_{<}-n_>=n(x)[2\textrm{Re}\{\textrm{Erf}[\sqrt{\ln2-x^{2}/2}]\}-1]
\end{equation}
If the correlation $c$ between longitudinal spin and motional energy is not
$100$\%, but say $c=60$\%, this result is reduced by the corresponding factor.
The dashed line in fig. \ref{mzx} is a plot of $0.6(n_<-n_>)$ according to (11);
a comparison with the other curves shows
that this model already gives a reasonable understanding
of the profile.

In conclusion, two major features of the new phenomenon are: a
cumulative effect of the ISRE producing a strong correlation between
motional energies and longitudinal spin directions, and its
stability over long times. The nature of the effect is more
ballistic than hydrodynamic. The shape of the profile depends on the
properties of the trap, not of the interactions: any phase space
distribution of the internal states that depends only on the energy
remains invariant under time evolution in the trap. ISRE therefore
plays a crucial role in creating the profile, but not in maintaining
it.

\emph{Acknowledgements. --} We thank J.E. Thomas and X. Du for
sharing their data with us and A. Amaricci for help with the
numerics.


\begin{thebibliography}{9}

\bibitem{Bashkin}E.P. Bashkin, JETP Lett. \textbf{33}, 8 (1981); JETP \textbf{60}, 1122 (1984).

\bibitem {Laloe}C.\ Lhuillier and F.\ Lalo\"{e}, J.\ Physique \textbf{43}, 197 (1982); \emph{ibid} \textbf{43}, 225 (1982).

\bibitem {H}B.R.\ Johnson, J.S.\ Denker, N.\ Bigelow, L.P.\ Levy, J.H.\ Freed
and D.M.\ Lee, Phys.\ Rev.\ Lett. \textbf{52}, 1508 (1984).

\bibitem {He}P.J. Nacher, G. Tastevin, M. Leduc, S.B. Crampton and F. Lalo\"{e}, J. Phys. Lett. \textbf{\ 45}, L-441 (1984); \textbf{\ 46}, 249 (1985).

\bibitem {Gully}W.J.\ Gully and W.J.\ Mullin, Phys.\ Rev.\ Lett \textbf{52},
1810 (1984).

\bibitem {Cornell}H.J.\ Lewandowski, D.M.\ Harber, D.L.\ Whitaker and
E.A.\ Cornell, Phys. Rev. Lett. \textbf{88}, 070403 (2002).

\bibitem {Levitov} M. \"O. Oktel and L.S. Levitov, Phys. Rev. Lett. \textbf{88}, 230403 (2002).

\bibitem{Fuchs}J.N. Fuchs, D.M. Gangardt, and F. Lalo\"e, Phys. Rev. Lett. \textbf{88}, 230404 (2002); Eur. Phys. J. D \textbf{25}, 57 (2003).

\bibitem{Williams}J.E. Williams, T. Nikuni, and C.W. Clark, Phys. Rev. Lett. \textbf{88}, 230405 (2002).


\bibitem{Thomas}X. Du, L. Luo, B. Clancy, and J.E. Thomas, Phys. Rev. Lett. \textbf{101}, 150401 (2008).

\bibitem{Bigelow} N.P. Bigelow, J.H Freed and D.M. Lee, Phys. Rev. Lett. \textbf{63}, 1609 (1989).

\bibitem{inpreparation}{F. Pi\'echon and J.N. Fuchs, in preparation.}



\end{thebibliography}
\end{document}